\documentclass{svproc}
\def\be{\begin{equation}}
\def\ee{\end{equation}}

\usepackage{url}
\usepackage{amssymb}

\usepackage{graphicx}% Include figure files

\usepackage{bm}% bold math

\begin{document}
\mainmatter              % start of a contribution
\title{Gauging Flavor Symmetries of the Standard Model: from Dark Energy to Matter - Antimatter Asymmetry from Higher Dimensions in the Early Universe.}
\titlerunning{Matter - Antimatter Asymmetry from Flavor Symmetry in Higher Dimensions}  % abbreviated title (for running head)
\author{ Anupam Singh }
\authorrunning{Anupam Singh} % abbreviated author list (for running head)
%\tocauthor{Anupam Singh}
%
\institute{ Physics Department, LNMIIT, Jaipur, India. \\
\email{ singh@lnmiit.ac.in }}

%\\ WWW home page:
%\texttt{https://www.lnmiit.ac.in/Department/%PhysicsPhysics_FacultyProfile.aspxnDeptID=76}}

\maketitle              % typeset the title of the contribution

\begin{abstract}
The Standard Model of Elementary Particle Physics has a global $U_1(1) \times U_2(1) \times U_3(1)$ flavor symmetry (often called an accidental symmetry because it was not put in by hand). Here the subscripts $i = 1,2,3$ refer to the indices corresponding to the 3 families in the Standard Model. It has previously been shown that the breaking of these symmetries at low energy may result in producing Dark Energy which is the dominant component of the energy density of the Universe. 
It has also previously been shown that this model of Dark Energy not only explains the accelerated expansion of the universe but has additional observational consequences consistent with observations and also makes predictions which may be verified in the near future.
Thus, this model of Dark Energy allows a space and time dependent Dark Energy as indicated by recent observations.
Further, in this model of Dark Energy the collapse of space dependent Dark Energy configurations can lead to the formation of Dark Energy Black Holes in both the Supermassive Black Hole category and the Intermediate Mass Black Hole category leading to predictions that may soon be verified by observations.  
Here we wish to examine the potential implication of these symmetries at high energies and in the Early Universe. In particular, we consider what might happen if these global symmetries get gauged at higher energies and that the global symmetries today are just a consequence of the gauge fields acquiring a constant vacuum expectation value below some energy scale $v$. Since, we have not yet seen signatures of such gauge fields at any of our particle accelerators, this would imply that $v$ is greater than the electroweak scale. Thus, in the early Universe at the time when the $U_i(1)$ gauge fields would be dynamical, the Higgs would not yet have acquired a non-zero vacuum expectation value and all Standard Model particles would be massless. The gauging of these global symmetries is desirable from a high energy physics and quantum gravity perspective. The chiral dynamics has a natural interpretation in terms of D Branes and our 4-dimensional Universe embedded in higher dimensions. Anomaly cancellation in this scenario results from the sum of the cubic terms vanishing as described below.  An important point to note is that the analysis and discussion in this article can provide an explanation for one of the long standing mysteries of the natural world: the matter - antimatter asymmetry observed in our Universe.
% We would like to encourage you to list your keywords within
% the abstract section using the \keywords{...} command.
\end{abstract}
\section{Introduction}

In the last century, a large fraction of fundamental physics was uncovered as a result of precision experiments carried out at high energy accelarators on earth. During the same period, humankind also started focusing on information on fundamental physics that could be obtained from astronomical observations. There is now a wide range of sophisticated satellites gathering data from the sky which is giving us fundamental new insights and opening unexpected doors to new physics.
A prime example of this is  the accurate determination of the Hubble constant and
the independent determination of the age of the universe which forced us to
critically re-examine the simplest cosmological model
- a universe with a zero cosmological constant\cite{Pierce,Freedman}.
These observations led us\cite{Singh} to examine the idea of
a non-vanishing vacuum energy due to fields as playing an important role in the current Universe.
Since then, there has been a growing body of work
that has confirmed the existence of what we now call dark energy. For an exhaustive exposition on Dark Energy, please see\cite{MarkAlessandra}.

Since Dark Energy forms by far the most dominant component of the Universe, we expect it to play an important role in our Universe. Indeed, it has been previously shown that Dark Energy can result in many observable consequences such as the formation of Dark Energy Black Holes\cite{DarkEnergyCollapseAndBHs,FiniteTemperatureFlavorSymmetries,FieldTheoryAndDarkEnergyUpdate}, Quasars\cite{Quasars} and Gravitational Waves\cite{DarkEnergyGravitationalWaves,DEGW_IceAges}. It has been found that symmetry and the dynamics of symmetry breaking as a result of phase transitions\cite{nonequilibrium,HolSing,cmupitt,us1,usfrw,rockymarkanupam,chiralpt,chiralpt2} has played a major role in the observable consequences of Dark Energy.

In a recent paper \cite{FiniteTemperatureFlavorSymmetries}, we have shown that the $U_1(1) \times U_2(1) \times U_3(1)$ flavor symmetry of the Standard Model can be responsible for the Dark Energy which is today the dominant component of the energy density of our Universe. Thus, it is clear that the $U_1(1) \times U_2(1) \times U_3(1)$ flavor symmetry of the Standard Model can play a very important role at low energies in our current late stage of the evolution of the Universe. For an extensive discussion of the $U_1(1) \times U_2(1) \times U_3(1)$ flavor symmetry of the Standard Model and its phenomenological implications please see \cite{SMsymmetries}.

At this point, I would like to emphasize 2 things about the $U_1(1) \times U_2(1) \times U_3(1)$ flavor symmetries being discussed here. First, these $U_1(1) \times U_2(1) \times U_3(1)$ flavor symmetries are an integral part of the Standard Model of elementary particle physics and both their existence and their breaking by small neutrino masses are consistent with all known phenomena in nature as previously discussed in detail for example in \cite{SMsymmetries}. Second, the breaking of these symmetries at low energies by non-vanishing neutrino masses has been extensively discussed earlier and can lead to many interesting phenomenomenological cosequences observed in nature including dark energy\cite{FiniteTemperatureFlavorSymmetries}, Black Holes\cite{DarkEnergyCollapseAndBHs,FiniteTemperatureFlavorSymmetries}, Quasars\cite{Quasars} and Gravitational Waves\cite{DarkEnergyGravitationalWaves,DEGW_IceAges}.

As previously discussed, this model of Dark Energy leads to a space and time dependent Dark  Energy\cite{FiniteTemperatureFlavorSymmetries} as indicated by recent observations from the Dark Energy Survey (DES) \cite{DES} and the Dark Energy Spectroscopic Instrument (DESI) \cite{DESI}.

Here we wish to focus on implications of the $U_1(1) \times U_2(1) \times U_3(1)$ flavor symmetry of the Standard Model at high energies and in the Early Universe.

\section{Gauging the Global Symmetries}

We first note that the  $U_1(1) \times U_2(1) \times U_3(1)$ flavor symmetry of the Standard Model are global symmetries at the low energies that have been explored so far.
Before going on to an exploration involving the gauging of these global symmetries we briefly mention our initial motivation for doing this and by the time we are done we will note some very interesting consequences of consistently gauging these global symmetries.

First, let us discuss the initial motivations. There have been long standing concerns about the effects of quantum gravity on global symmetries - please see for example\cite{GlobalSymmetriesAndGravity}. By gauging these global symmetries we can ensure that in reality and at higher energies we are only dealing with gauge symmetries.

Another motivation for gauging these global symmetries arises from D Brane physics and trying to connect it to models of elementary particle physics - please see for example\cite{LeontarisEtAl} and references therein.
It is well known that when one has N coincident D Branes we get a $U(N)$ gauge theory whereas all of the models of elementary particles starting with the electroweak theory to QCD  and extensions of the standard model such as the Pati-Salam model, $SU(5)$ GUTs etc. all involve $SU(N)$ gauge theories. Every time, we use N coincident Branes to give us an SU(N) gauge theory, we are left with an "extra" $U(1)$ gauge symmetry. One might be left wondering where all the extra $U(1)$ gauge symmetries went. Of course,we do have global $U(1)$ symmetries in the Standard Model as already noted above. Thus, if we can gauge the global $U(1)$ symmetries present at low energies in the Standard Model, then the mystery of the missing $U(1)$'s may be solved and we can smoothly match the low energy physics to the high energy physics as discussed below.

\section{Symmetries, Anomalies and Anomaly Cancellation}

The $U_i(1)$ flavor symmetries of the Standard Model that we are discussing are also called the Lepton Family symmetries $L_i$ for each of the 3 families in the Standard Model labeled by the 3 family indices $ i = 1,2,3$. The electron family has the index $i = 1$, the muon family has the index $i = 2$ and tau family has the index $i = 3$.

The corresponding symmetry currents are given by 
\begin{equation}
 j_{i \mu} = \bar{\psi_i} \gamma_\mu \psi_i
\end{equation}
and the axial or chiral current is given by 
\begin{equation}
j_{i \mu}^{5} = \bar{\psi_i} \gamma_\mu \gamma_5 \psi_i
\end{equation}

for $i = 1,2,3$ corresponding to the 3 families.

Conservation of the symmetry current $j_{i \mu}$ is given by the equation $\partial^\mu j_{i \mu} = 0$ and gives us the non-zero conserved charges $Q_i$ as per the Standard Model:

\begin{center}
\begin{tabular}{ |c|c|c|c| } %{Table 1: Charges}
 \hline
 \multicolumn{4}{|c|}{\bf{Table 1: Charges $Q_i$}} \\
\hline
 \bf{Particles} & $Q_1$ & $Q_2$ & $Q_3$ \\ 
\hline
 Electron & +1 & 0 & 0 \\
\hline 
 Positron & --1 & 0 & 0\\ 
\hline
 Electron Neutrino & +1 & 0 & 0 \\
\hline 
 Electron Antineutrino & --1 & 0 & 0\\
 \hline
 Muon & 0 & +1 & 0 \\
\hline 
 Anti-Muon & 0 & --1 & 0\\ 
\hline
 Muon Neutrino & 0 &  +1& 0 \\
\hline 
 Muon Antineutrino &0  &  --1& 0\\
\hline
Tau & 0 &0 & +1  \\
\hline 
Anti-Tau & 0 &  0 & --1\\ 
\hline
Tau Neutrino & 0 & 0& +1 \\
\hline 
Tau Antineutrino &0 &0 & --1\\

 \hline
\end{tabular}
\end{center}

Please note that for all other particles the charges $Q_i = 0$.

At this point, we first note the ABJ anomaly equation for the axial current:
\begin{equation}
\partial^{\mu}j_{i \mu}^{5} = - \frac{Q_i^2}{8 \pi^2} F_i^{\mu \nu} F^*_{i \mu \nu}
\end{equation}
where $F^*_{i \mu \nu} $ is the dual field strength tensor given by  
\begin{equation}
F^*_{i \mu \nu} = \frac{1}{2} \epsilon_{\mu \nu \lambda \rho} F_i^{\lambda \rho}
\end{equation}
Please note that the index $i$ is the family index and there is no sum over $i$ implied in the anomaly equation above.

Of course, in order to gauge the global flavor symmetries $U_i(1)$ we have to introduce the Abelian gauge fields $A_{i \mu}$ where $i$ is the family index and $\mu$ is the space-time index. As always, the field strength tensors $F_{i \mu \nu}$ in this Abelian case is given by:
$ F_{i \mu \nu} = \partial_\mu A_{i \nu}    - \partial_\nu A_{ i \mu} $ .

It should be noted that in this case, since we have $U_i(1)$ symmetries involved, we essentially have 3 copies of elctromagnetism labeled by the family index $i = 1,2,3$ except that the charges for each such copy of electromagnetism is different and given by the Table 1 above listing the $Q_i$ corresponding to each copy of electromagnetism labeled by $i$. This makes determining the dynamics straightforward.

First, we note that in order to gauge the $U_i(1)$ symmetries, the ABJ anomaly does not need to vanish\cite{Anomalies} - it is only the gauge anomaly which is proportional to the sum over the $Q_i^3$ which needs to vanish. Thus, in the case we are interested in, since the charges come in $+1,-1$ pairs or are $0$ (please see Table 1 above), the sum  over the  $Q_i^3$ actually vanishes and hence we can gauge the  the $U_i(1)$ symmetries as there in no obstruction from anomalies. 

Thus, we can consistently build a picture where at low energies the flavor symmetries are global symmetries but at high energies these flavor symmetries become gauge symmetries.
Essentially, this means that above some high energy scale which we will denote by $v$, the gauge fields are dynamical and can have a non-trivial space-time dependence. However, below the energy scale $v$, these gauge fields acquire their uniform space independent vacuum expectation values. We further note that the scale $v$ must be above the electroweak symmetry breaking scale because otherwise we would have seen the dynamical signature of these extra gauge fields at existing particle accelerators such as the LHC.

We now turn our focus towards connecting our discussion so far with the observable aspects of particle physics and cosmology.

For this purpsoe we first look at the Dirac equation and its potential interpretation in terms of D-Branes.

\section{The Dirac Equation and D-Branes}

The Dirac equation can be written in the form:

\begin{equation}
	\label{eqn: Dirac}
	\left( i \gamma^\mu \partial_\mu - m \right) \psi = 0 
\end{equation}
%\label{Dirac}

The discussion that follows is conveniently carried out in terms of the states $\psi_{L}$ and $\psi_{R}$, where we have introduced the left-handed projections: $\psi_{L} = (1 - \gamma^5) \psi/2 $ and right-handed projection: $\psi_{R} = (1 + \gamma^5) \psi/2 $.

First, we note that in the massless $m = 0$ limit the 2 states $\psi_L$ and $\psi_R$ decouple and are independent of each other.

Moreover, we note that the $\psi_L$ states are positive energy solutions to the Dirac Equation and represent particle states or matter and the $\psi_R$ states are negative energy solutions and represent anti-particle states or anti-matter.

It is a fact of nature that it is only the $\psi_L$ states which interact with the electroweak gauge bosons. Furthermore, in the D Brane picture of the Standard Model - please see  e.g. the work of Leontaris et. al. -\cite{LeontarisEtAl} -  the electroweak gauge bosons and $\psi_L$ are confined to the 2 coincident branes giving rise to the $SU(2)$ gauge fields of the Standard Model. After the Higgs field acquires a non-zero vacuum expectation value, the mass term connects the $\psi_L$ to the $\psi_R$ - this happens after the electroweak symmetry breaking.

At this point, we note that all matter particles carry $ +1 $ charge and are located on $L$ Branes all anti-matter particle (anti-particles) carry $ -1$ charge and are located on $R$ Branes.

The discussion above leads to the point of view that we live on the Left-handed Brane on which all the matter particles that we detect as well as the electroweak gauge bosons live. This statement may raise a concern about the fact that we do occasionally come across anti-matter particles. But this concern is immediately alleviated by noting that the appearance of the anti-matter particles is determined precisely by the Dirac equation (\ref{eqn: Dirac}) given above. Thus, from this perspective, the anti-matter particles are visitors to our Left-handed Brane and their appearance is strictly regulated by the Dirac equation. Since the Dirac equation has been verified a very large number of times and there is no known violation of this equation this puts the point of view put forth here on a firm phenomenological footing.

Realistic D-Brane models explicitly showing the role of the extra  $U_i(1)$ symmetries in the context of the Standard Model and its extensions such as the Pati-Salam model are extensively discussed by Anastasopoulos et.al.\cite{U1DBranesAndSMextensions}

\section{Brane Collisions and the Matter-Antimatter Asymmetry creation in the Early Universe}

As we raise the temperature of our Universe and effectively travel into it's past, first we cross the electroweak phase transition - at higher energy densisties - the Higgs vacuum expectation value is zero and all Standard Model particles are massless at this point. At the electroweak phase transition, the flavor symmetries are still global - otherwise we would have seen the signature of the extra gauge bosons at existing accelerators. However, at some higher energy scale $v$ the flavor symmetries can become dynamical gauge fields. At this point, since the flavor symmetries are carried by $U_i(1)$ gauge fields, we have extra copies of electromagnetic type forces - except the charges are not the usual electromagnetic charges but instead given by the charges as already  enumerated in Table 1 above.

Further, we re-iterate that all matter particles carry $ +1 $ charge and are located on $L$ Branes all anti-matter particle (anti-particles) carry $ -1$ charge and are located on $R$ Branes.

Thus, even if the $L$ and $R$ were located close together in the early universe as one might expect, the $L$ - $R$ symmetry can be easily broken by an additional Brane approaching or colliding with the $L$ and $R$ Brane pair initially located close to each other.

This will not only enable the $L$ - $R$ symmetry breaking needed for the Pati-Salam\cite{PatiSalam} or any other $L$-$R$ symmetric extension\cite{LeftRightSymmetry}  of the Standard Model but also provide an explanation for the Matter-Antimatter Asymmetry which follows from this higher demensional perspective on the Early Universe as described here.

\section{Summary and Conclusions}

It has been recently realized\cite{FiniteTemperatureFlavorSymmetries}, that the $U_1(1) \times U_2(1) \times U_3(1)$ flavor symmetry of the Standard Model can be responsible for the Dark Energy which is today the dominant component of the energy density of our Universe. Thus, it is clear that the $U_1(1) \times U_2(1) \times U_3(1)$ flavor symmetry of the Standard Model can play a very important role at low energies in our current late stage of the evolution of the Universe. For an extensive discussion of the $U_1(1) \times U_2(1) \times U_3(1)$ flavor symmetry of the Standard Model and its phenomenological implications please see \cite{SMsymmetries}.

Since  the $U_1(1) \times U_2(1) \times U_3(1)$ flavor symmetry of the Standard Model plays such an important role in our Universe it behooves us to examine the various implications of these flavor symmetries.

We started off by noting that the  $U_1(1) \times U_2(1) \times U_3(1)$ flavor symmetry of the Standard Model are global symmetries at the low energies that have been experimentally explored so far. We further noted that  there have been long standing concerns about the effects of quantum gravity on global symmetries\cite{GlobalSymmetriesAndGravity}. 
By gauging these global symmetries we can ensure that in reality and at higher energies we are only dealing with gauge symmetries.
Further motivation for gauging these global symmetries arises from D Brane physics and trying to connect it to models of elementary particle physics\cite{LeontarisEtAl}.
It is well known that when one has N coincident D Branes we get a $U(N)$ gauge theory whereas all of the models of elementary particles starting with the Standard Model and extensions of the standard model such as the Pati-Salam model, $SU(5)$ GUTs etc. all involve $SU(N)$ gauge theories. Every time, we use N coincident Branes to give us an SU(N) gauge theory, we are left with an "extra" $U(1)$ gauge symmetry. Thus, if we can gauge the global $U(1)$ symmetries present at low energies in the Standard Model, then the mystery of the missing $U(1)$'s may be solved and we can smoothly match the low energy physics to the high energy physics. Indeed, the discussion above may be viewed as a bottom-up phenomenological approach proceeding from low energies to higher energies. This may eventually contribute to understanding the full higher dimensional physics and dynamics in a manner which enables obtaining our observed universe in a low energy limit.

Motivated by these considerations, we examined the consequences of gauging the global flavor symmetries of the Standard Model and this has paid us rich dividends as illustrated above.

We see from the perspective presented here that matter and anti-matter are living on 2 different surfaces and that this may provide a basis for the observed matter-antimatter symmetry in our Universe. Furthermore, this approach also naturally provides a mechanism for Left-Right symmetry breaking in models such as the Pati-Salam model or any Left-Right symmeric extension of the Standard Model.

\section{Acknowledgements}

For stimulating and helpful discussions I would like to thank Prof. A. P. Balachandran.

%\begin{figure}
%\vspace{2.5cm}
%\caption{This is the caption of the figure}
%\end{figure}

%\section{Summary and Discussion}


\end{document}